\documentclass[final,3p,12pt]{elsarticle}
\usepackage{amssymb}
\usepackage{amsmath}
\newcommand{\bm}{\mathbf}

\journal{Journal of Computational Physics}
\begin{document}
\begin{frontmatter}
\title{Analysis of optical waveguides
with arbitrary index profile using an immersed interface method}
\author{Theodoros P. Horikis}
\address{Department of Mathematics, University of Ioannina, Ioannina 45110, Greece}
\begin{abstract}
A numerical technique is described that can efficiently compute solutions in
interface problems. These are problems with data, such as the coefficients of
differential equations, discontinuous or even singular across one or more
interfaces.  A prime example of these problems are optical waveguides and as
such the scheme is applied to Maxwell's equations as they are formulated to
describe light confinement in Bragg fibers. It is based on standard finite
differences appropriately modified to take into account all possible
discontinuities across the waveguide's interfaces due to the change of the
refractive index. Second and fourth order schemes are described with additional
adaptations to handle matrix eigenvalue problems, demanding geometries and
defects.
\end{abstract}

\end{frontmatter}
Light confinement due to cylindrical Bragg reflection instead of total internal
reflection was first proposed more than three decades ago \cite{yeh} and gave
birth to the so-called Bragg fibers. These fibers attract considerable interest
because of their ability to guide light in an air core; they are essentially
dielectric coaxial fibers comprised of alternating circular layers with
different indices of refraction. The key to making these fibers confine light
efficiently, i.e., have low absorption loss and a high threshold power for
nonlinear effects, is to use materials with a high index contrast
\cite{fink,nature,science2,johnson}. However, the high index contrast and the
layered structure that gives these fibers their unique properties also makes
them difficult to model.

Mathematically, these problems are called interface problems since their input
data (such as the coefficients of differential equations, source terms etc.)
may be discontinuous or even singular across one or several interfaces. The
solution to an interface problem, therefore, typically is non-smooth or even
discontinuous across the interfaces. Interface problems occur in many physical
applications, particularly for free boundary/moving interface problems, such
as, the modeling of the Stefan problem of solidification process and crystal
growth, composite materials, multi-phase flows, cell and bubble deformation,
and many others \cite{li_review}.

Several methods have been proposed to study these problems, including
asymptotic analysis \cite{xu.ol}, the transfer matrix method \cite{guo2},
finite element methods \cite{eguchi,eguchi2}, special functions --Bessel
\cite{kawanishi} and Hankel \cite{sakai}-- formalism, and Galerkin numerical
methods \cite{guo,barai}. A comparative analysis of the most commonly used
methods has also been published \cite{guo2}, demonstrating the capabilities and
limitations of each method. Among the different numerical solution methods, the
finite difference (FD) method is more attractive due to its advantage of simple
formulation and numerical implementation and thus will be used here to analyze
these problem.

Our approach is based on the immersed interface method \cite{leveque} (IIM)
which has been attracting considerable attention due to the many physical
problems that can be applied on
\cite{zhong,berth,xu,khoo,karagiozis,horikis.pla,costinett,deng,rutka}. This
approach has two additional advantages over standard Galerkin methods
\cite{guo}. First of all, the scheme does not need to be modified significantly
if different boundary conditions are used, thus allowing to calculate all
possible solutions without any modifications. Methods based upon the Galerkin
method typically require a set of basis functions that naturally satisfy the
boundary conditions, hence the solution must be reformulated in a significant
way if these change. More importantly, the IIM does not depend on any specific
functional representation of solutions. Hence, cumbersome integrations or
finding roots of nontrivial functions, such as Bessel functions (even when
asymptotically approximated \cite{xu.ol}), are avoided.

The essence of the method is to appropriately modify the correct matrix
elements of a standard (central) FD scheme so as to take into account all
discontinuities across interfaces. Starting with a differential equation and
under a FD approximation one transforms the equation into an algebraic system
of the form
\[
A\bm{x}=\bm{b}
\]
where $\bm{x}$ is the solution or
\[
A\bm{x}=\lambda\bm{x}
\]
for eigenvalue problems, where $\lambda$ is the eigenvalue. The matrix $A$ is
comprised of zero elements except for the main, upper and lower diagonal (for
second order accurate solutions or more for more accurate schemes), i.e.
\[
A = \left( {\begin{array}{ccccc}
  {{a_{11}}}&{{a_{12}}}&0&\cdots&0 \\
  {{a_{21}}}&{{a_{22}}}&{{a_{23}}}&\cdots&0 \\
  0&{{a_{32}}}&{{a_{33}}}&{{a_{34}}}&\cdots \\
  {0}&{0}&{\ddots}&{\ddots}&{\ddots}
\end{array}} \right)
\]
This feature gives the FD method an additional advantage since $A$ is
tridiagonal. Using sparse matrix algebra one can significantly lower
computational time, whether the inverse of the matrix or the eigenvalues is
shout for. The goal is to identify the elements where the interface occurs and
correct the appropriate matrix elements in a way described below to take into
account the effects of the interfaces. Remarkably, these corrections are
solutions of linear algebraic systems of equations. Thus, the matrix remains
sparse and computational time is kept to a minimum.

The original formulation \cite{leveque} of the IIM does not consider eigenvalue
problems such as the problems of interest here. Hence, in order to deal with
waveguide problems for Bragg fibers, the method must be extended to handle any
eigenvalue problem described by a second order differential operator. Moreover,
the method must be extended to handle coupled equations like the ones
describing the two polarization components of the electromagnetic field.
Furthermore, since the method is based on finite differences one can use higher
order schemes to increase the accuracy of the calculations. The extension to
higher order accuracy is also presented in this article. However, extending to
higher order posses a major limitation. In some problems, the geometry of the
interfaces are such that in order to have enough points between them (using a
uniform grid) requires to increase the total number of points and as such
computational time. To overcome this, we introduce a coordinate stretching
transformation which allows the method to handle these more demanding
geometries.

The article is organized as follows: We begin with the description of the
method in second order. In so doing, we extend the original formulation of the
IIM to matrix eigenvalue problems. In addition, it is shown that all
discontinuities/singularities are removed from the equation and passed on the
FD scheme as corrections to the standard FD coefficients based on matching
conditions across an interface. These corrections are calculated using linear
systems of algebraic equations. Then fibers with deformations are considered to
further illustrate the versatility of the method. Finally, we extend to fourth
order and conclude with more demanding geometries in which the original IIM
would fail unless a coordinate stretching is applied.
\section{Formulation}
The vector Helmholtz equations in cylindrical coordinates for the magnetic
field are \cite{snyder}
\begin{subequations}
\begin{gather}
  \nabla_t^2 H_r  - \frac{2}
{{r^2 }}\frac{{\partial H_\theta  }} {{\partial \theta }} - \frac{1}
{{r^2 }}H_r  + k^2 n^2 H_r  = \beta ^2 H_r \\
  \nabla_t^2 H_\theta   - \frac{1}
{r}\frac{{d(\ln n^2 )}} {{dr}}\frac{\partial } {{\partial r}}(rH_\theta ) +
\frac{1} {r}\left[ {\frac{{d(\ln n^2 )}} {{dr}} + \frac{2} {r}}
\right]\frac{{\partial H_r }} {{\partial \theta }} - \frac{1} {{r^2 }}H_\theta
+ k^2 n^2 H_\theta   = \beta ^2 H_\theta \\
H_z=\frac{i}{\beta}\frac{1}{r}\left[ \frac{\partial}{\partial r}(rH_r) +
\frac{\partial H_\theta}{\partial\theta} \right] \label{helm33}
\end{gather}
\label{helm}
\end{subequations}
where $k$ is the wavenumber, $n=n(r)$ is the (arbitrary) index of refraction,
and $\beta$ is the propagation constant. We focus on the first two equations,
since the components of the electrical field, $E_r$, $E_\theta$ and $E_z$, as
well as the transverse magnetic field, $H_z$ can be recovered from $H_r$ and
$H_\theta$ using Maxwell's equations. All fields of the guiding modes are
assumed to go to zero as $r\rightarrow\infty$. In addition, in order for system
(\ref{helm}) to be well defined at the origin, the following boundary
conditions must hold at $r=0$
\begin{subequations}
\begin{gather}
\frac{\partial^2 H_\theta}{\partial\theta^2}-H_\theta +2\frac{\partial
H_r}{\partial \theta}=0 \\
\frac{\partial^2 H_r}{\partial\theta^2}-H_r -2\frac{\partial
H_\theta}{\partial \theta}=0
\end{gather}%
\label{helm.bc}
\end{subequations}
Separation of variables in Eqs. (\ref{helm}) suggests that the fields can be
expressed in the form $H_r(r,\theta )= H_{rm}(r)\cos(m\theta)$ and
$H_\theta(r,\theta )= H_{\theta m}(r)\sin(m\theta)$ with $m$ an integer. Hence,
Eqs.~(\ref{helm}) become
\begin{subequations}
\begin{gather}
\frac{1} {r}\frac{d} {{dr}}\left( {r\frac{{dH_{rm} }} {{dr}}} \right) -
\frac{1} {{r^2 }}\left[ {(1 + m^2 )H_{rm}  + 2mH_{\theta m} } \right] + k^2n^2
H_{rm}  = \beta ^2 H_{rm} \label{helm2.1}
\\
\frac{n^2} {r}\frac{d} {{dr}}\left[ {\frac{1} {{n^2 }}\left(
{r\frac{{dH_{\theta m} }} {{dr}} + H_{\theta m}  + mH_{rm} } \right)} \right]
-\frac{1} {{r^2 }}\left[ {2mH_{rm}  + (1 + m^2 )H_{\theta m} } \right]
\nonumber \\ \hspace{4.5cm} - \frac{1} {r}\frac{d} {{dr}}\left( {mH_{rm}  +
H_{\theta m} } \right) + k^2n^2 H_{\theta m}  = \beta ^2 H_{\theta m}
\label{helm2.2}
\end{gather}
\label{helm2}
\end{subequations}
In the absence of angular dependence, i.e., $m=0$, these equations uncouple and
describe the TE and TM modes of the fiber, respectively. Hereafter we drop the
double subscript notation and we set $H_{rm}=H_r(r)$ and $H_{\theta
m}=H_\theta(r)$. The boundary conditions as $r\rightarrow\infty$ remain the
same and at $r=0$ Eqs. (\ref{helm.bc}) become
\begin{subequations}
\begin{gather}
(1+m^2)H_r+2mH_\theta=0 \\ 2mH_r+(1+m^2)H_\theta=0
\end{gather}%
\label{helm2.bc}%
\end{subequations}%
When $m{\neq}1$ these simply imply that $H_r(r=0)=H_\theta(r=0)=0$. When
$m{=}1$, however, Eqs.~(\ref{helm2.bc}) are identical and an additional
boundary condition must be imposed. It is straightforward to show via a Taylor
series expansion around $r=0$ that the solution will satisfy the condition
\[
\frac{d}{dr}(H_r-H_\theta)=0 \label{m1.bc}
\]
which we will impose as our additional boundary condition. The way to implement
these boundary conditions into the FD scheme is shown in the appendix.
\section{The second order method}\label{iim.coupled}
Consider the system of coupled equations that describe the electric, $H_r$, and
magnetic, $H_\theta$, fields in a circular waveguide, Eqs. (\ref{helm2}).
Expanding all derivatives in Eqs. \eqref{helm2} and after appropriate
simplifications the system is written as
\begin{subequations}
\begin{gather}
  \frac{d^2H_r}{dr^2}  + \frac{1}
{r}\frac{dH_r}{dr}  - \frac{{2m}} {{r^2 }}H_\theta   + \left( {k^2 n^2  -
\frac{{m^2  + 1}}
{{r^2 }}} \right)H_r  = \beta ^2 H_r  \\
  \frac{d^2H_\theta}{dr^2}   + \left( { - \frac{{2n'}}
{n} + \frac{1} {r}} \right)\frac{dH_\theta}{dr}  - m\left( {\frac{{2n'}} {{nr}}
+ \frac{2} {{r^2 }}} \right)H_r  + \left( {k^2 n^2  - \frac{{m^2  + 1}} {{r^2
}} - \frac{{2n'}} {{nr}}} \right)H_\theta   = \beta ^2 H_\theta
\end{gather}\label{helm3}
\end{subequations}
and the prime ($'$) denotes differentiation with respect to $r$. Consider a
finite difference approximation for Eqs. (\ref{helm3}) of the form (central
differences)
\begin{subequations}
\begin{gather}
  \gamma _1 H_{r,i - 1}  + \gamma _2 H_{r,i}  + \gamma _3 H_{r,i + 1}  + \Delta H_{\theta ,i}  = \beta ^2 H_{r,i}\\
  \delta _1 H_{\theta ,i - 1}  + \delta _2 H_{\theta ,i}  + \delta _3 H_{\theta ,i + 1}  + \Gamma H_{r,i}  = \beta ^2
  H_{\theta ,i}
\end{gather}
\label{te.new}
\end{subequations}
where (see appendix)
\[
\gamma _1 =\delta_1 = \frac{1} {{h^2 }} - \frac{1} {{2hr_i}},\;\;\gamma _2=\delta_2  =  - \frac{2}
{{h^2 }} + k^2 n^2  - \frac{1} {{r_i^2 }},\;\;\gamma _3 =\delta_3 = \frac{1} {{h^2 }} +
\frac{1} {{2hr_i}},\;\; \Gamma=\Delta=-\frac{2m}{r_i^2}
\]
with $r\in[a,b]$ defined on a uniform grid as
\[
r_i=a+ih=a+i\left(\frac{b-a}{N}\right), \quad i=0,1,2,...,N.
\]
Using this one can find the correct row where the correction must be applied.
If the interface occurs at $r=r^*$ then setting
\[
r_i=r^* \Rightarrow j=\text{int}\left\{\frac{r^*-a}{h}\right\}
\]
gives the row to be corrected. The function int$\{\}$ denotes integer part.

In Eq. (\ref{helm3}) the index of refraction is a discontinuous function and
changes, say, at $r = r^*$, so that
\[
n(r) = \left\{ {\begin{array}{c}
   {n_1 ,\quad r < r^* }  \\
   {n_2 ,\quad r > r^* }
 \end{array} } \right.
\]
The above scheme cannot be used without any modifications as it does not take
into consideration the singularities appearing in the equations due to the form
of the index of refraction. Thus we reformulate the problem, including the
differential equations, in vector form. In addition, the terms including
derivatives of discontinuous functions are neglected (we assume the index is
piecewise constant) and their contribution is incorporated into the finite
difference scheme through appropriate jump conditions on the interfaces. In so
doing, Eqs. (\ref{helm3}) read in vector form
\begin{gather}
{\bm{H}}_{rr}  + \frac{1} {r}{\bm{H}}_r  + B{\bm{H}} = \beta^2
{\bm{H}}\label{helm.vec}
\end{gather}
where $\bm{H}=(H_r,H_\theta)^\mathrm{T}$ and
\[
B = \left( {\begin{array}{cc}
   k^2 n^2  - (m^2  + 1)/r^2 & - 2m/r^2 \\
   -2m/r^2 & k^2 n^2  - (m^2  + 1)/r^2
 \end{array} } \right)
\]
In the vector formulation subscripts denote differentiation.

\begin{figure}[!htbp]
    \centering
    \includegraphics[width=4in]{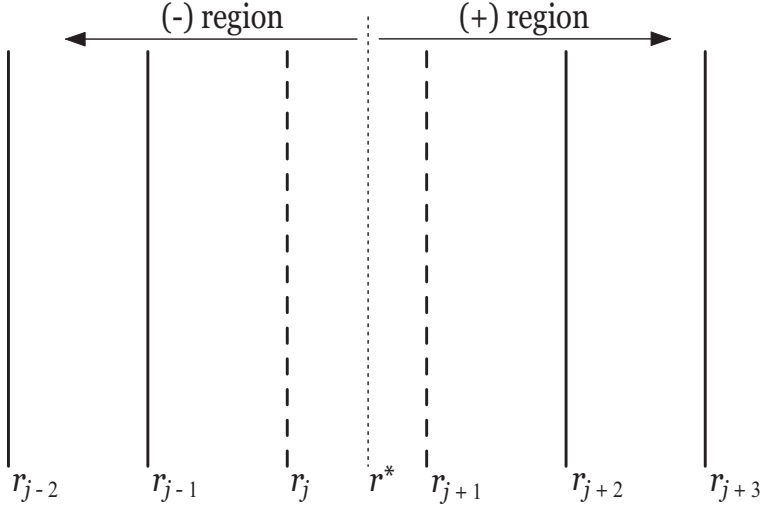}
    \caption{The $(-)$ and $(+)$ regions, the problematic point
    $r=r^*$ and the irregular grid points $r_j$ and $r_{j+1}$.}
    \label{index.fig}
\end{figure}

The goal is to determine the coefficients of the finite difference
approximation (\ref{te.new}) to take into account this jump of the refractive
index at $r = r^*$. To do this, divide the region $[a,b]$ into two, the $(-)$
region for $r < r^*$ an the $(+)$ region for $r > r^*$, as in Fig.
\ref{index.fig}.
The analysis is similar for the two regions, but needs to be repeated for both.
Start with the $(-)$ region: we need to replace $\bm{H}_{i-1}$, $\bm{H}_i$,
$\bm{H}_{i+1}$ in Eqs. (\ref{te.new}) so that the local truncation error is
first order. Expand $\bm{H}$ around the points before and after the jump,
namely
\begin{subequations}
\begin{eqnarray}
  \bm{H}(r_{j - 1} )  = \bm{H}_{j - 1}  &=& \bm{H}^ -   + (r_{j - 1}  - r^* )\bm{H}_r^ -   + \frac{1}
{2}(r_{j - 1}  - r^* )^2\bm{H}_{rr}^ -   \\
  \bm{H}(r_j ) = \bm{H}_j  &=& \bm{H}^ -   + (r_j  - r^* )\bm{H}_r^ -   + \frac{1}
{2}(r_j  - r^* )^2 \bm{H}_{rr}^ -   \\
  \bm{H}(r_{j + 1} )  = \bm{H}_{j + 1}  &=& \bm{H}^ +   + (r_{j + 1}  - r^* )\bm{H}_r^ +   + \frac{1}
{2}(r_{j + 1}  - r^* )^2 \bm{H}_{rr}^ +  \label{h3.taylor}
\end{eqnarray}\label{h.taylor}
\end{subequations}
The index $j$ denotes points closest to the jump, as in Fig. \ref{index.fig}.
Notice that we only include second order terms in the expansions. We need to
replace the $(+)$ functions in Eq. (\ref{h3.taylor}) since we are in the $(-)$
region; this is done through the continuity conditions.

To derive these continuity or matching conditions one needs to refer to the
physical properties of the problem. Here all fields are continuous functions
across all interfaces, i.e.
\begin{eqnarray*}
H_r^+ &=& H_r^- \\
H_\theta^+ &=& H_\theta^-\\
H_z^+ &=& H_z^-
\end{eqnarray*}
The last equation and Eq. \eqref{helm33} also yield (recall that all fields are
only functions of $r$)
\[
\frac{i}{\beta}\frac{1}{r^*}\left(H_r^+ r^*\frac{d}{dr}H_r^+ \right) =
\frac{i}{\beta}\frac{1}{r^*}\left(H_r^- r^*\frac{d}{dr}H_r^- \right)
\]
or
\[
\frac{d}{dr}H_r^+ = \frac{d}{dr}H_r^-
\]
Another condition may be derived by integrating Eq. \eqref{helm2.2} around
$r=r^*$, namely
\begin{gather*}
\lim_{\Delta r\rightarrow 0}\int_{r^*-\Delta r}^{r^*+\Delta r}\left\{\frac{n^2}
{r}\frac{d} {{dr}}\left[ {\frac{1} {{n^2 }}\left( {r\frac{{dH_{\theta} }}
{{dr}} + H_{\theta }  + mH_{r} } \right)} \right] -\frac{1} {{r^2 }}\left[
{2mH_{r}  + (1 + m^2 )H_{\theta} } \right]\right\} \nonumber \\ - \lim_{\Delta
r\rightarrow 0}\int_{r^*-\Delta r}^{r^*+\Delta r}\left\{ \frac{1} {r}\frac{d}
{{dr}}\left( {mH_{r}  + H_{\theta} } \right) + k^2n^2 H_{\theta}\right\}  =
\lim_{\Delta r\rightarrow 0}\int_{r^*-\Delta r}^{r^*+\Delta r}\left\{ \beta ^2
H_{\theta} \right\}
\end{gather*}
and since all fields are continuous across $r=r^*$ the only nonzero remaining
terms are
\begin{gather*}
\frac{1}{n_2^2}\left( r^* \frac{dH_\theta^+}{dr} + H_\theta^+ +m H_r^+ \right)-
(mH_r^+ +H_\theta^+) =\\
\frac{1}{n_1^2}\left( r^* \frac{dH_\theta^-}{dr} + H_\theta^- +m H_r^- \right)-
(mH_r^- +H_\theta^-)
\end{gather*}
Finally, the matching conditions for the second derivatives are a consequence
of the continuity of the fields and Eqs. \eqref{helm3} since
\begin{gather*}
\bm{H}^-=\bm{H}^+ \Leftrightarrow \beta^2\bm{H}^-=\beta^2\bm{H}^+
\end{gather*}
which results in
\[
\bm{H}_{rr}^++\frac{1}{r^*}\bm{H}_r^++B^+\bm{H}^+=\bm{H}_{rr}^-+\frac{1}{r^*}\bm{H}_r^-+B^-\bm{H}^-
\]
In summary, the continuity conditions in vector form are
\begin{subequations}
\begin{eqnarray}
\bm{H}^ - &=& \bm{H}^ + \label{cont.vec1}\\
\bm{H}_r^ +   &=& C \bm{H}_r^ -   + D \bm{H}\label{cont.vec2}\\
\bm{H}_{rr}^ +   &=& \bm{H}_{rr}^ -   + E \bm{H}_r^ -   + F \bm{H}
\label{cont.vec3}
\end{eqnarray}\label{cont.vec}
\end{subequations}
where
\begin{gather*}
  C = \left( \begin{array}{cc}
   1 & 0  \\
   0 & n_2^2/n_1^2
 \end{array} \right),\quad
 D = \frac{n_2^2 /n_1^2  - 1}
{r^* }\left( \begin{array}{cc}
   0 & 0  \\
   m & 1
 \end{array} \right),\quad
  E =  - \frac{n_2^2 /n_1^2  - 1}
{r^* }\left( \begin{array}{cc}
   0 & 0  \\
   0 & 1
 \end{array}\right),\\
 F =  - \left( \begin{array}{cc}
   k(n_2^2  - n_1^2 ) & 0  \\
   m(n_2^2 /n_1^2  - 1)/r^{*2} & (n_2^2 /n_1^2  - 1)/r^{*2} + k(n_2^2  - n_1^2 )
 \end{array} \right)
\end{gather*}
To put everything together return to the FD approximation in matrix form
\[
\Gamma _1 {\bm{H}}_{i - 1}  + \Gamma _2 {\bm{H}}_i  + \Gamma _3
{\bm{H}}_{i + 1}  = \beta^2 {\bm{H}}_i
\]
where the scalar coefficients  $\gamma$'s are replaced by $2 \times 2$
matrices. Replacing $\bm{H}_{j-1}$, $\bm{H}_j$ and $\bm{H}_{j+1}$ from Eqs.
\eqref{h.taylor}, using Eqs. \eqref{cont.vec} and
\[
\beta^2 \bm{H}_j=\bm{H}_{rr}^- +\frac{1}{r^*}\bm{H}_r^- +B^-\bm{H}
\]
we obtain an equality with $\bm{H}_{rr}^-$, $\bm{H}_r^-$ and $\bm{H}$ on both
sides. Matching the relative coefficients results in the following linear
system for the coefficients at $r_j\leq r^*$
\begin{gather*}
  \Gamma _1  + \Gamma _2  + \Gamma _3 \left[ {I_2  + (r_{j + 1}  - r^* )D + \frac{1}
{2}(r_{j + 1}  - r^* )^2 F} \right] = B^ -   \\
  (r_{j - 1}  - r^* )\Gamma _1  + (r_j  - r^* )\Gamma _2  + \Gamma _3 \left[ {(r_{j + 1}  - r^* )C + \frac{1}
{2}(r_{j + 1}  - r^* )^2 E} \right] = \frac{1}
{{r^* }}I_2  \\
  \frac{1}
{2}(r_{j - 1}  - r^* )^2 \Gamma _1  + \frac{1} {2}(r_j  - r^* )^2 \Gamma _2  +
\frac{1} {2}(r_{j + 1}  - r^* )^2 \Gamma _3  = I_2
\end{gather*}
and at $r_{j+1}>r^*$
\begin{gather*}
  \Gamma _1 \left[ {I_2  - (r_j  - r^* )C^{ - 1} D + \frac{1}
{2}(r_j  - r^* )^2 F_2 } \right] + \Gamma _2  + \Gamma _3  = B^ +   \\
  \Gamma _1 \left[ {(r_j  - r^* )C^{ - 1}  + \frac{1}
{2}(r_j  - r^* )^2 E_2 } \right] + (r_{j + 1}  - r^* )\Gamma _2  + \Gamma _3
(r_{j + 2}  - r^* ) = \frac{1}
{{r^* }}I_2  \\
  \frac{1}
{2}(r_j  - r^* )^2 \Gamma _1  + \frac{1} {2}(r_{j + 1}  - r^* )^2 \Gamma _2  +
\frac{1} {2}(r_{j + 2}  - r^* )^2 \Gamma _3  = I_2
\end{gather*}
where we need to introduce the matrices
\begin{gather*}
E_2  = \left( {\begin{array}{cc}
   0 & 0  \\
   0 & (1 - n_1^2 /n_2^2)/r^*
 \end{array} } \right),\\
 F_2  = \left( {\begin{array}{cc}
   {k^2 (n_2^2  - n_1^2 )} & 0  \\
   {m({1 - n_1^2 /n_2^2 })/
{r^* }} & {k^2 (n_2^2  - n_1^2 ) - ({1 - n_1^2 /n_2^2 })/ {r^{*2} }}
 \end{array} } \right)
\end{gather*}
For the latter system (the $(+)$ side, $r_{j+1}>r^*$) Eqs. \eqref{cont.vec}
were inverted to substitute for the $(-)$ side. Each of the above systems
represents a $12\times 12$ system of algebraic equations that determines the
coefficients of the matrices. If multiple interfaces are present, one merely
applies these difference formulas multiple times.  Note that the result is a
system of finite difference equations each involving three neighboring points
making the resulting equations tridiagonal. Because of the tridiagonal
structure of the matrix, sparse matrix algebra can be used to determine the
eigenvalues and eigenmodes. Thus, a large number of points can be used for
modest computational cost, which allows the accuracy of the results to be
increased and the modes of complicated structures to be determined, e.g., Bragg
fibers with many thin layers \cite{johnson}. Also note that the corrections
depend only on the values of the refractive index before and after the
discontinuity.  This means that the jump conditions do not have to be modified
if the index varies radially between the discontinuities.

\begin{figure}[!htbp]
    \centering
    \includegraphics[width=4in]{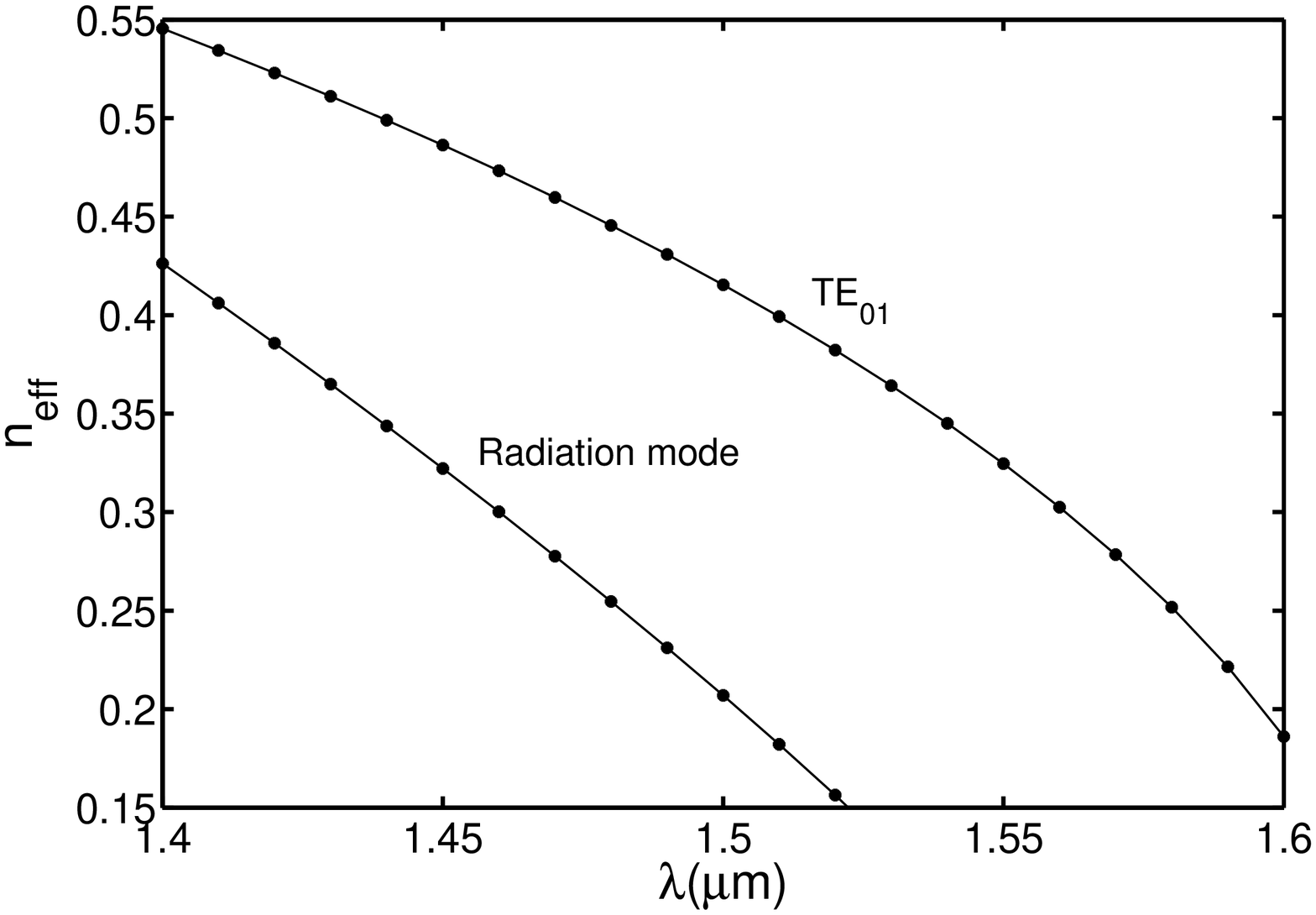}
    \includegraphics[width=4in]{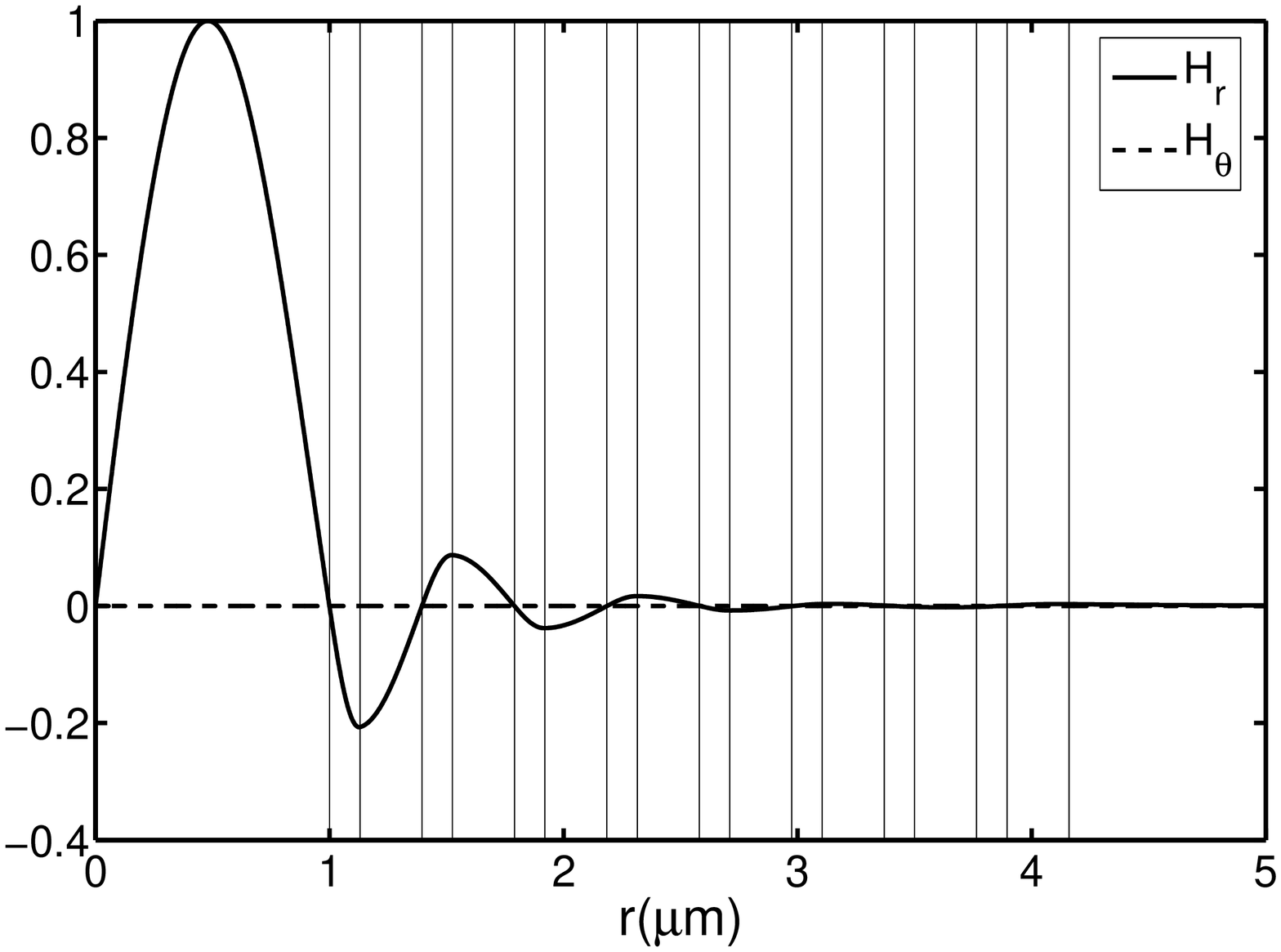}
    \includegraphics[width=4in]{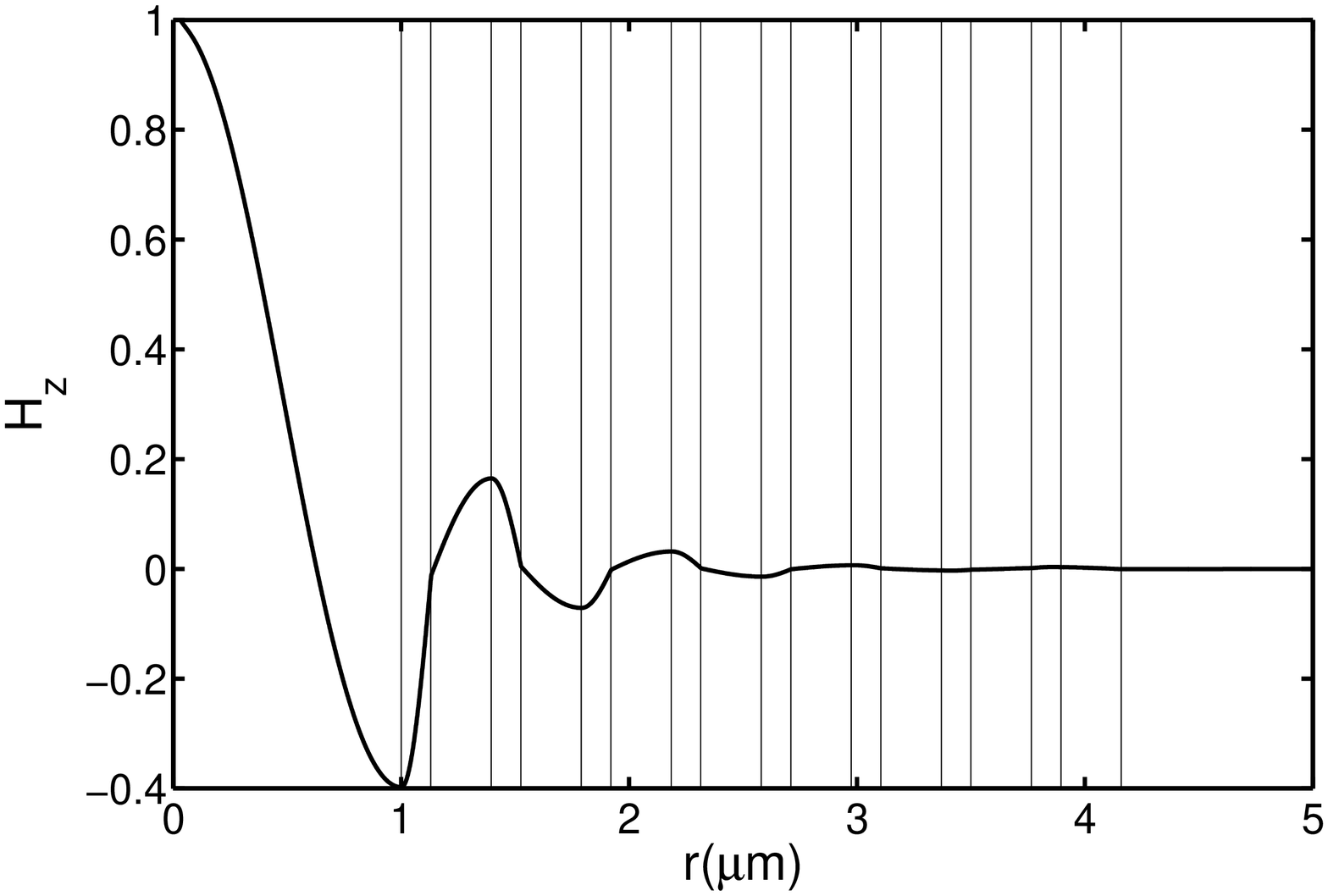}
    \caption{Dispersion (top) of the fundamental TE mode in the air-core Bragg fiber
    described in the text and in Refs. \cite{xu.ol,guo} and the TE, TM (middle) and $H_z$ (bottom) modes plotted at
    $\lambda=1.55\,\mu$m.}
    \label{disp.fig}
\end{figure}

To test our method we use a Bragg fiber with an air core of radius $1.0\,\mu$m
and a cladding that consists of alternating layers with refractive indices
$n_1=3.0$ and $n_2=1.5$ \cite{xu.ol,guo}. The distance between layers is
$0.130\,\mu$m and $0.265\,\mu$m. As done previously \cite{guo}, an imaginary
cladding with a refractive index close to zero is added outside the core to
prevent reflections. For multilayer Bragg fibers the effective index,
$\beta/k$, is usually measured instead of just $\beta$. Within the spectral
range of $1.4\,\mu m < \lambda < 1.6\,\mu m$, the Bragg fiber supports a single
TE mode, whose propagation constant effective index is plotted in Fig.
\ref{disp.fig}.

\begin{figure}[!htbp]
    \centering
    \includegraphics[width=4in]{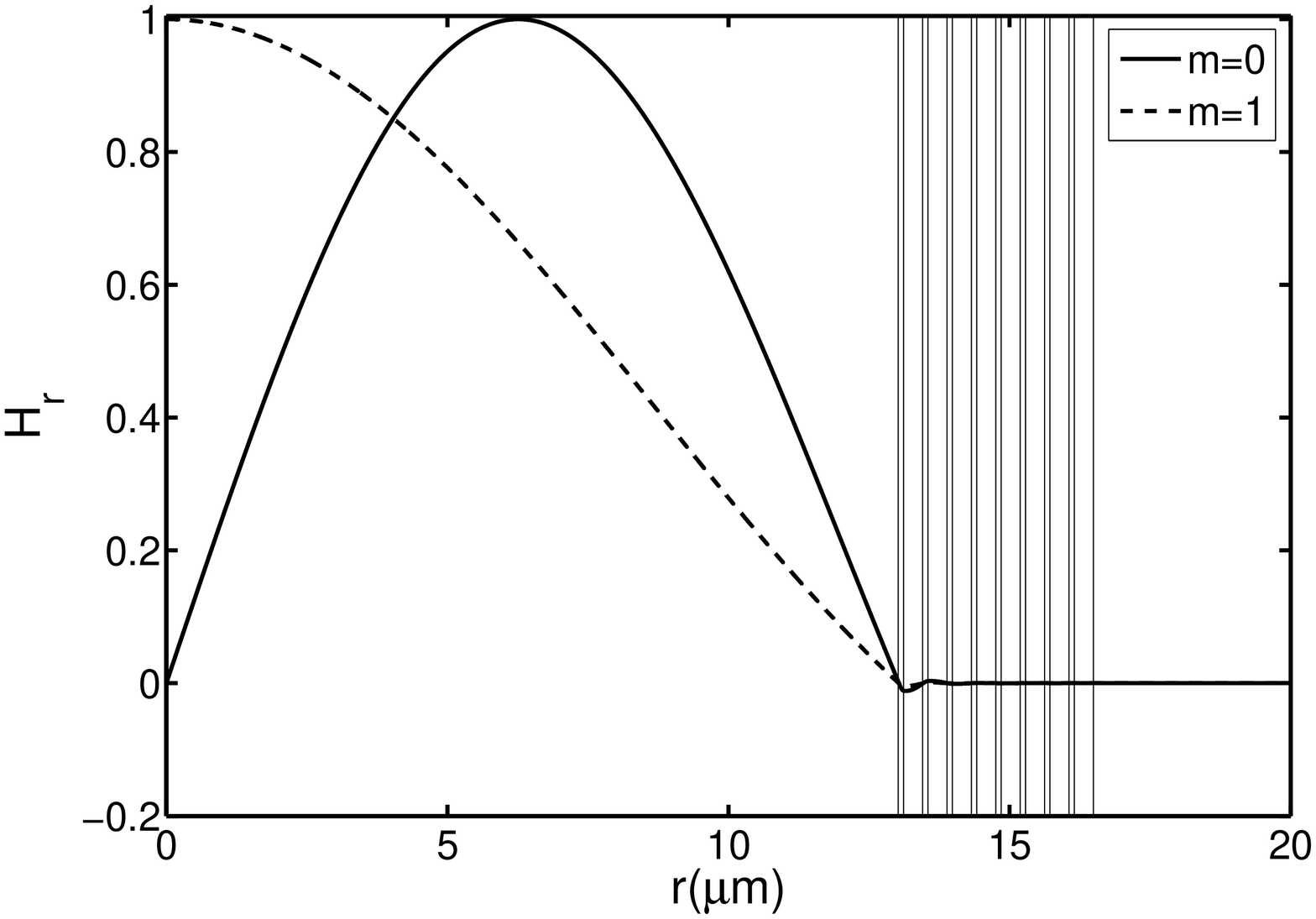}
    \includegraphics[width=4in]{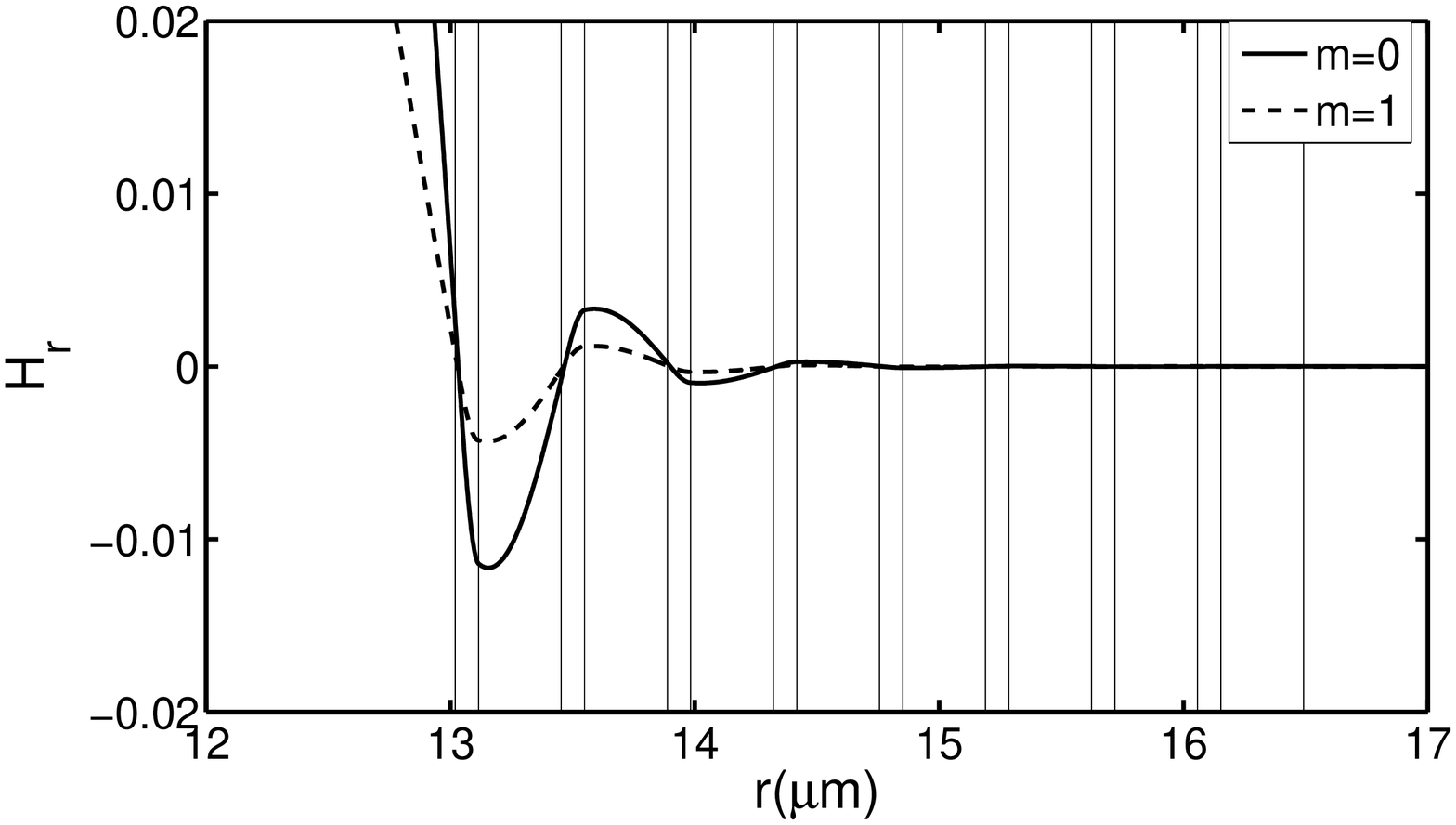}
    \caption{The fundamental TE mode and the $TE_{11}$ mode of the air-core Omniguide fiber
    described in the text and in Ref. \cite{johnson}. The bottom figure is a blow-up of the fields
    near the end of the fiber's core.}
    \label{omni1.fig}
\end{figure}

This method can also be applied to more complicated and computationally
demanding fibers. For example, let's consider the Omniguide fiber described in
Ref. \cite{johnson}. This is a large air-core fiber with core radius
$13.02\mu$m surrounded by $17$ layers, starting with a high-index layer, with
indices $n_1=4.6$ and $n_2=1.6$, and thicknesses $l_1=0.09548\mu$m and
$l_2=0.33852\mu$m, respectively. Under these parameters the wavelength for the
lowest dissipation losses is $\lambda=1.55\mu$m. The first two modes of the
fiber, namely the $TE_{01}$ and $TE_{11}$ are plotted in Fig. \ref{omni1.fig}.
The effective indices are $0.99736632$ and $1.00097643$, respectively. The way
to implement the additional boundary condition for the $TE_{11}$ mode ($m=1$)
is described in the appendix.

Notably, plastic optical fibers (POFs) which attracted recent attention because
of their use in subscriber line systems and home networks also have a large
core diameter and a high core-cladding refractive-index difference compared
with conventional silica glass multimode optical fibers and can support tens of
thousands to hundreds of thousands of propagation modes. A recent finite
element method was used \cite{eguchi,eguchi2} to analyze their properties. We
discuss these (and more demanding geometries) at a later section.
\section{Fibers with deformations}
Single defects surrounded by Bragg reflectors as the basis for annular
resonators were proposed and analyzed in Ref. \cite{scheuer}. The basic
geometry was a circumferentially-guiding defect is located within a medium
which consists of annular Bragg layers. As a result of the circular geometry,
the layer widths, unlike in rectangular geometry, are not constant, and the
task is to determine the widths that lead to maximum confinement in the defect.
In addition, it has been suggested \cite{skorobogatiy} that fibers with such
defects can be used to model pairs of identical touching hollow Bragg fibers.
The dielectric profile along the interfiber center line resembles a
one-dimensional Bragg grating with a central defect formed by the two external
layers of the fiber mirrors.

\begin{figure}[!htbp]
    \centering
    \includegraphics[width=4in]{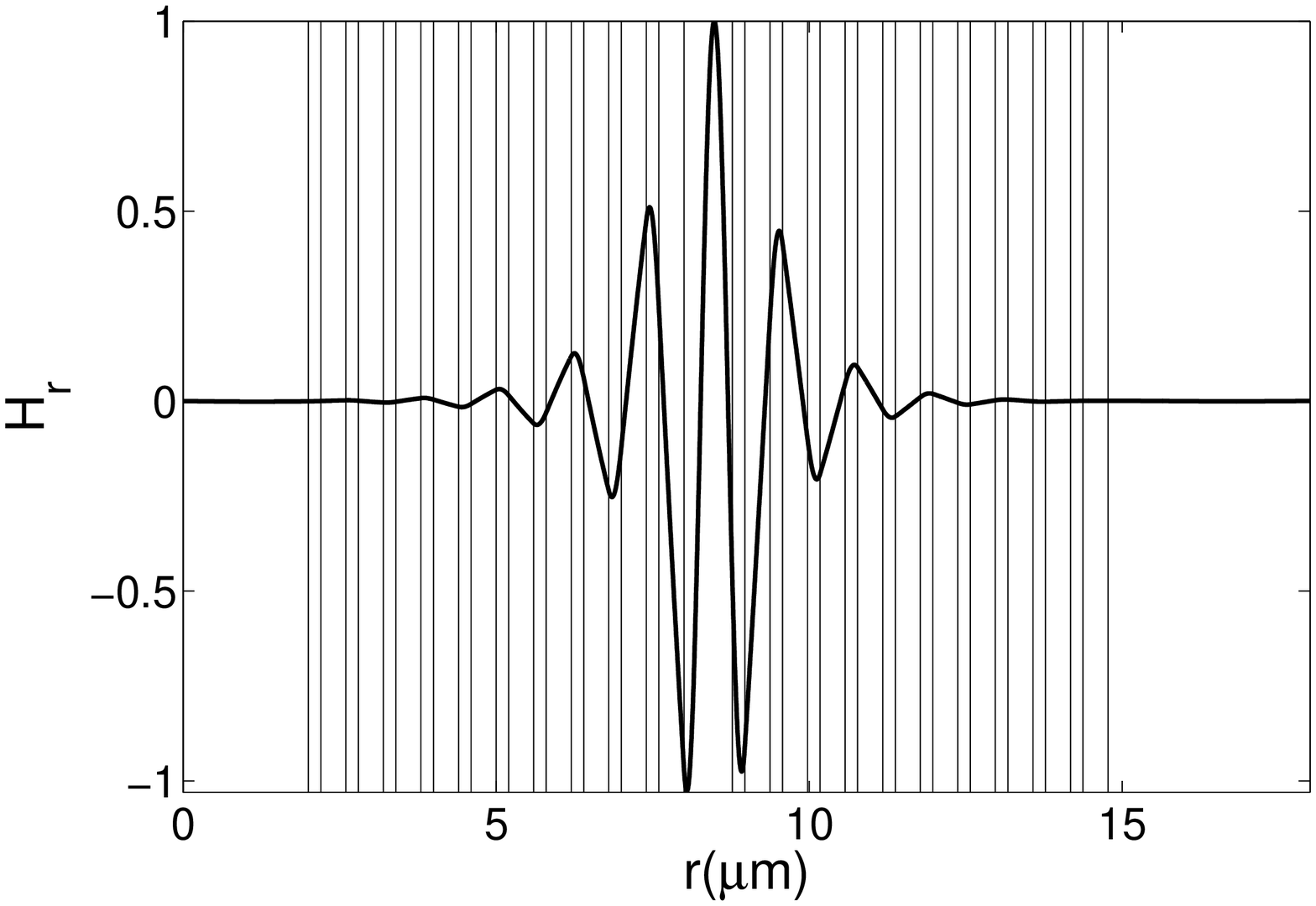}
    \caption{The magnetic field distribution of an annular defect mode resonator.}
    \label{defect1.fig}
\end{figure}

Figure \ref{defect1.fig} depicts the magnetic field inside a defect. The high
index layers and the defect have an effective refractive index $n_1=2$ while
the low index layers have an effective refractive index $n_2=1$. The internal
and external Bragg reflectors have $10$ periods, and the wavelength is
$1.45\mu$m. The defect is $(\lambda/2)\mu$m wide. The effective index is found
to be $0.92257830$.

Resonant features correspond to the points of accidental degeneracy of
$TE_{01}$ with higher-order modes. This is true for this case and the effective
indices for the $TE_{01}$ and $TE_{20}$ modes where found to be $0.87238117$
and $0.49188825$, respectively. For more details on the physical properties of
these defects we refer the reader to \cite{skorobogatiy}.
\section{Higher order method}
Higher order accuracy methods can also be derived in a similar manner. The main
focus is again Eqs. (\ref{helm2}). As before, assume the refractive index to be
piece-wise constant, and write these equations in vector form as
\begin{gather}
{\bm{H}}_{rr}  + \frac{1} {r}{\bm{H}}_r  + B_1 {\bm{H}} = \beta ^2
{\bm{H}} \label{helm.h4}
\end{gather}
where $\bm{H}=(H_r,H_\theta)^\mathrm{T}$,
\[
B_1  = \left( \begin{array}{cc}
   k^2 n^2  - (m^2  + 1)/r^2& -2m/r^2\\
   -2m/r^2 & k^2 n^2  - (m^2  + 1)/r^2
 \end{array} \right)
 \]
and the subscripts denote differentiation. Recall, the continuity conditions
are
\begin{subequations}
\begin{eqnarray}
{\bm{H}}^ -   &=& {\bm{H}}^ +\\
{\bm{H}}_r^ +   &=& C_1 {\bm{H}}_r^ -   + D_1 {\bm{H}}\\
{\bm{H}}_{rr}^ +   &=& {\bm{H}}_{rr}^ -   + E_1 {\bm{H}}_r^ -   +
F_1 {\bm{H}}
\end{eqnarray}
\label{cont.h4}
\end{subequations}
where
\begin{gather*}
  C_1  = \left( {\begin{array}{cc}
   1 & 0  \\
   0 & n_2^2/n_1^2
 \end{array} } \right),\quad
 D_1  = \frac{{n_2^2 /n_1^2  - 1}}
{{r^* }}\left( {\begin{array}{cc}
   0 & 0  \\
   m & 1
 \end{array} } \right),\quad
  E_1  =  - \frac{{n_2^2 /n_1^2  - 1}}
{{r^* }}\left( {\begin{array}{cc}
   0 & 0  \\
   0 & 1
 \end{array} } \right),\\
 F_1  =  - \left( {\begin{array}{cc}
   {k(n_2^2  - n_1^2 )} & 0  \\
   m(n_2^2 /n_1^2  - 1)/r^{*2} & (n_2^2 /n_1^2  - 1)/r^{*2} + k(n_2^2  - n_1^2 )
 \end{array} } \right)
\end{gather*}
The fourth order finite difference approximation in matrix form is
\begin{gather}
\Gamma _1 {\bm{H}}_{i - 2}  + \Gamma _2 {\bm{H}}_{i - 1}  + \Gamma _3
{\bm{H}}_i  + \Gamma _4 {\bm{H}}_{i + 1}  + \Gamma _5 {\bm{H}}_{i
+ 2}  = \beta ^2 {\bm{H}}_i \label{fd.h4}
\end{gather}
where (see appendix)
\begin{eqnarray*}
  \Gamma _1  &=& \left(\begin{array}{cc}
   -1/12h^2+1/12r_i h & 0  \\
   0 & -1/12h^2 + 1/12r_i h
 \end{array}  \right),\\
 \Gamma _2  &=& \left( \begin{array}{cc}
   4/3h^2 - 2/3r_i h & 0  \\
   0 & 4/3h^2 - 2/3r_i h
 \end{array}  \right), \\
  \Gamma _3  &=& \left( \begin{array}{cc}
   k^2 n^2  -(m^2  + 1)/r_i^2 - 5/2h^2 & - 2m/r_i^2\\
   -2m/r_i^2& k^2 n^2  - (m^2  + 1)/r_i^2 - 5/2h^2
 \end{array}  \right), \\
  \Gamma _4  &=& \left( \begin{array}{cc}
   4/3h^2 + 2/3r_i h & 0  \\
   0 & 4/3h^2 + 2/3r_i h
 \end{array}  \right),\\
 \Gamma _5  &=& \left( \begin{array}{cc}
   1/12h^2 - 1/12r_i h & 0  \\
   0 & -1/12h^2 - 1/12r_i h
 \end{array}  \right)
\end{eqnarray*}
at all regular points. We need to define these matrices for the irregular
points using the IIM. We define our grid as before, namely $r_i  = a + i\frac{b
- a}{N}$, with $N$ the total number of points, and assume that the point
$r=r^*$ is between two points, say $r_j$ and $r_{j+1}$. As in the case of
second order accuracy, we define a point to be regular if all points of the
finite difference equation, Eq. (\ref{fd.h4}), are on the same region, either
the $(-)$ or the $(+)$ regions. All other points are irregular. Thus, in Fig.
\ref{index.fig} the irregular points are $r_{j - 1} ,r_j ,r_{j + 1}$ and
$r_{j+2}$. We only need to define the $\Gamma$'s in Eq. (\ref{fd.h4}) at these
points.

To do so we need to expand in Taylor series all function around the problematic
point $r=r^*$ up to and including terms of fourth order. Hence, for example at
$r=r_j<r^*$
\begin{eqnarray*}
{\bm{H}}_{j-1}  &=& {\bm{H}} + (r_{j-1}  - r^* ){\bm{H}}_r^ -   + \frac{1}
{2}(r_{j-1}  - r^* )^2 {\bm{H}}_{rr}^ -   + \frac{1} {6}(r_{j-1}  - r^* )^3
{\bm{H}}_{rrr}^ -   + \frac{1} {{24}}(r_{j-1}  - r^* )^4 {\bm{H}}_{rrrr}^-
\\
{\bm{H}}_j  &=& {\bm{H}} + (r_j  - r^* ){\bm{H}}_r^ -   + \frac{1}
{2}(r_j  - r^* )^2 {\bm{H}}_{rr}^ -   + \frac{1} {6}(r_j  - r^* )^3
{\bm{H}}_{rrr}^ -   + \frac{1} {{24}}(r_j  - r^* )^4 {\bm{H}}_{rrrr}^-
\\
{\bm{H}}_{j+1}  &=& {\bm{H}} + (r_{j+1}  - r^* ){\bm{H}}_r^ +   + \frac{1}
{2}(r_{j+1}  - r^* )^2 {\bm{H}}_{rr}^ +   + \frac{1} {6}(r_{j+1}  - r^* )^3
{\bm{H}}_{rrr}^ +   + \frac{1} {{24}}(r_{j+1}  - r^* )^4 {\bm{H}}_{rrrr}^+
\\
{\bm{H}}_{j+2}  &=& {\bm{H}} + (r_{j+2}  - r^* ){\bm{H}}_r^ +   + \frac{1}
{2}(r_{j+2}  - r^* )^2 {\bm{H}}_{rr}^ +   + \frac{1} {6}(r_{j+2}  - r^* )^3
{\bm{H}}_{rrr}^ +   + \frac{1} {{24}}(r_{j+2}  - r^* )^4 {\bm{H}}_{rrrr}^+
\end{eqnarray*}
While the continuity conditions are known up to the second derivative, see Eqs.
\eqref{cont.h4}, additional are needed for the higher derivatives. These are
obtained from differentiating Eq. \eqref{helm.vec} and using Eqs.
\eqref{cont.h4}. Hence for the third derivative
\begin{eqnarray*}
\bm{H}_{rrr}^+ +\frac{1}{r^*}\bm{H}_{rr}^+ + B_r \bm{H}_r^+=\beta^2 \bm{H}_r^+ =
\beta^2 (C_1 \bm{H}_r^- +D_1 \bm{H})
\end{eqnarray*}
which after further use of Eqs. \eqref{helm.vec} and \eqref{cont.h4} becomes
\[
{\bm{H}}_{rrr}^ +   = C_3 {\bm{H}}_{rrr}^ -   + D_3 {\bm{H}}_{rr}^ -   + E_3
{\bm{H}}_r^ -   + F_3 {\bm{H}}
\]
and similarly for the fourth derivative
\[
{\bm{H}}_{rrrr}^ +   = {\bm{H}}_{rrrr}^ -   +
C_5 {\bm{H}}_{rrr}^ - + D_5 {\bm{H}}_{rr}^ -   + E_5 {\bm{H}}_r^ -   + F_5
{\bm{H}}
\]
where
\begin{eqnarray*}
  C_3  &=& \left( \begin{array}{cc}
   1 & 0  \\
   0 & n_2^2/n_1^2
 \end{array}  \right),\\
 D_3  &=& \left( \begin{array}{cc}
   0 & 0  \\
   m(n_2^2 /n_1^2  - 1)/r^*& 2(n_2^2 /n_1^2  - 1)/r^*
 \end{array} \right), \\
  C_5  &=& \left( \begin{array}{cc}
   0 & 0  \\
   0 & - 2(n_2^2 /n_1^2  - 1)/r^*
 \end{array}  \right),\\
 D_5  &=& \left( \begin{array}{cc}
   - 2k^2 (n_2^2  - n_1^2 ) & 0  \\
   - 2m(n_2^2 /n_1^2  - 1)/r^{*2} & - 2k^2 (n_2^2  - n_1^2 ) - 4(n_2^2 /n_1^2  - 1)/r^{*2}
 \end{array} \right), \\
  E_3  &=& \left( \begin{array}{cc}
   - k^2 (n_2^2  - n_1^2 ) &  - 2m^2(n_2^2 /n_1^2  - 1)/r^{*2} \\
   - m(n_2^2 /n_1^2  - 1)/r^{*2}& - k^2 (n_2^2  - n_1^2 )n_2^2/n_1^2 + 2(n_2^2
/n_1^2  - 1)/r^{*2}
 \end{array}  \right), \\
  E_5  &=& \left( \begin{array}{cc}
   0 & - 12m(n_2^2 /n_1^2  - 1)/r^{*3}  \\
   2m(n_2^2 /n_1^2  - 1)/r^{*3}& 2k^2(n_2^2 /n_1^2  - 1)(n_2^2  - n_1^2 )/r^* - (4m^2 +
10)(n_2^2 /n_1^2  - 1)/r^{*3}
 \end{array}  \right)
\end{eqnarray*}
The elements of the other two matrices are
\begin{eqnarray*}
[F_3]_{11} &=& {\frac{{k^2 }} {{r^* }}(n_2^2  - n_1^2 ) + 2m^2 \frac{{n_2^2 /n_1^2
- 1}} {{r^{*3} }}}
\\
\left[F_3\right]_{12} &=& {2m^2 \frac{{n_2^2 /n_1^2  - 1}} {{r^{*3} }}}
\\
\left[F_3\right]_{21}&=&{ - m\frac{{n_2^2 /n_1^2  - 1}} {{r^* }}k^2 (n_2^2  - n_1^2 ) + 4m^2
\frac{{n_2^2 /n_1^2  - 1}} {{r^{*3} }}}
\\
\left[F_3\right]_{22}&=&{ - \frac{{k^2 }} {{r^*
}}\frac{{n_2^2 }} {{n_1^2 }}(n_2^2  - n_1^2 ) + 4\frac{{n_2^2 /n_1^2  - 1}}
{{r^{*3} }}}
\\
\left[F_5\right]_{11}&=&{k^4 (n_2^2  - n_1^2 )^2  - \frac{{3k^2 }}
{{r^{*2} }}(n_2^2  - n_1^2 ) - 12m^2 \frac{{n_2^2 /n_1^2  - 1}} {{r^{*4} }}}
\\
\left[F_5\right]_{12}&=&{ - 12m\frac{{n_2^2 /n_1^2  - 1}} {{r^{*4} }}}
\\
\left[F_5\right]_{21}&=& 2m\frac{{n_2^2 /n_1^2  - 1}} {{r^{*2} }}k^2 (n_2^2  - n_1^2 )^2  -
2(2m^2  + 7)m\frac{{n_2^2 /n_1^2  - 1}} {{r^{*4} }}
\\
\left[F_5\right]_{22}&=&  k^4 (n_2^2  - n_1^2 )^2  + \frac{{2n_2^2 /n_1^2  - 5}} {{r^{*2}
}}k^2 (n_2^2  - n_1^2 ) - (8m^2  + 14)\frac{{n_2^2 /n_1^2  - 1}} {{r^{*4} }} +
4m^2 \frac{{n_2^2 /n_1^2  - 1}} {{r^{*4} }}
\end{eqnarray*}

Since higher derivative are present we must complete the system of equations by
expanding the right hand side as follows
\begin{eqnarray*}
\beta ^2 {\bm{H}}_j  &=& \beta ^2 \left( {{\bm{H}} + (r_j  - r^*
){\bm{H}}_r^ -   + \frac{1}
{2}(r_j  - r^* )^2 {\bm{H}}_{rr}^ -  } \right) \\
   &=& \underbrace{{\bm{H}} + \frac{1}
{{r^* }}{\bm{H}}_r^ -   + B_1 {\bm{H}}_{rr}^ -}_{\beta^2 \bm{H}}   + (r_j  - r^* )\underbrace{\left(
{{\bm{H}}_{rrr}^ -   + \frac{1}
{{r^* }}{\bm{H}}_{rr}^ -   + B_2 {\bm{H}}_r^ -   + B_3 {\bm{H}}} \right)}_{\beta^2\bm{H}_r^-} \\
   &+& \frac{1}
{2}(r_j  - r^* )^2 \underbrace{\left( {{\bm{H}}_{rrrr}^ -   + \frac{1} {{r^*
}}{\bm{H}}_{rrr}^ -   + B_4 {\bm{H}}_{rr}^ -   + B_5 {\bm{H}}_r^ -
+ B_6 {\bm{H}}} \right)}_{\beta^2\bm{H}_{rr}^-}
\end{eqnarray*}
where we have differentiated Eq. (\ref{helm.h4}) twice and used the appropriate
continuity conditions. Also,
\begin{eqnarray*}
  B_2  &=& \left( \begin{array}{cc}
   k^2 n^2  - (m^2  + 2)/r^{*2}& -2m/r^{*2}\\
   -2m/r^{*2}& k^2 n^2  - (m^2  + 2)/r^{*2}
 \end{array}  \right),\\
 B_3  &=& \left( \begin{array}{cc}
   2(m^2  + 1)/r^{*3} & 4m/r^{*3}\\
   4m/r^{*3} & 2(m^2  + 1)/r^{*3}
 \end{array}  \right),\\
  B_4  &=& \left( \begin{array}{cc}
   k^2 n^2  - (m^2  + 3)/r^{*3}& -2m/r^{*3}\\
   -2m/r^{*3} & k^2 n^2  - (m^2  + 3)/r^{*3}
 \end{array} \right),\\
 B_5  &=& \left( \begin{array}{cc}
   (4m^2  + 6)/r^{*3}& 8m/r^{*3}\\
   8m/r^{*3}& (4m^2  + 6)/r^{*3}
 \end{array} \right),\\
 B_6  &=& \left( \begin{array}{cc}
   - 6(m^2  + 1)/r^{*4}& -12m/r^{*4}\\
   -12m/r^{*4} & - 6(m^2  + 1)/r^{*4}
 \end{array} \right)
\end{eqnarray*}
Now, we can form a system of algebraic equations to determine all the unknown
coefficients. This is done as before, by equating the coefficients of
$\bm{H}_{rrrr}^-$, $\bm{H}_{rrr}^-$, $\bm{H}_{rr}^-$, $\bm{H}_r^-$ and $\bm{H}$
at $r<r^*$ and their $(+)$ counterparts at $r>r^*$. Thus at $r=r_j<r^*$
\begin{gather*}
  {\Gamma _1} + {\Gamma _2} + {\Gamma _3} + {S_1}(j + 1){\Gamma _4} + {S_1}(j + 2){\Gamma _5} = {B_1} + ({r_j} -
  {r^*}){B_3} + \frac{1}
{2}{({r_j} - {r^*})^2}{B_6} \\
  ({r_{j - 2}} - {r^*}){\Gamma _1} + ({r_{j - 1}} - {r^*}){\Gamma _2} + ({r_j} - {r^*}){\Gamma _3} + {S_2}(j +
  1){\Gamma _4} + {S_2}(j + 2){\Gamma _5} =\\ \frac{1}
{{{r^*}}}{I_2} + ({r_j} - {r^*}){B_2} + \frac{1} {2}{({r_j} - {r^*})^2}{B_5}
\\
  \frac{1}
{2}{({r_{j - 2}} - {r^*})^2}{\Gamma _1} + \frac{1} {2}{({r_{j - 1}} -
{r^*})^2}{\Gamma _2} + \frac{1} {2}{({r_j} - {r^*})^2}{\Gamma _3} + {S_3}(j +
1){\Gamma _4} + {S_3}(j + 2){\Gamma _5} = \\{I_2} + \frac{{({r_j} - {r^*})}}
{{{r^*}}}{I_2} + \frac{1} {2}{({r_j} - {r^*})^2}{B_4} \\
  \frac{1}
{6}{({r_{j - 2}} - {r^*})^3}{\Gamma _1} + \frac{1} {6}{({r_{j - 1}} -
{r^*})^3}{\Gamma _2} + \frac{1} {6}{({r_j} - {r^*})^3}{\Gamma _3} + {S_4}(j +
1){\Gamma _4} + {S_4}(j + 2){\Gamma _5} = \\({r_j} - {r^*}){I_2} + \frac{1}
{{2{r^*}}}{({r_j} - {r^*})^2}{I_2} \\
  \frac{1}
{{24}}{({r_{j - 2}} - {r^*})^4}{\Gamma _1} + \frac{1} {{24}}{({r_{j - 1}} -
{r^*})^4}{\Gamma _2} + \frac{1} {{24}}{({r_j} - {r^*})^4}{\Gamma _3} + {S_5}(j
+ 1){\Gamma _4} + {S_5}(j + 2){\Gamma _5} =\\ \frac{1} {2}{({r_j} -
{r^*})^2}{I_2}
\end{gather*}
where ${S_1}(j) = {I_2} + ({r_j} - {r^*}){D_1} + \frac{1} {2}{({r_j} -
{r^*})^2}{F_1} + \frac{1} {6}{({r_j} - {r^*})^3}{F_3} + \frac{1} {{24}}{({r_j}
- {r^*})^4}{F_4}$, ${S_2}(j) = ({r_j} - {r^*}){C_1} + \frac{1} {2}{({r_j} -
{r^*})^2}{E_1} + \frac{1} {6}{({r_j} - {r^*})^3}{E_3} + \frac{1} {{24}}{({r_j}
- {r^*})^4}{E_5}$, ${S_3}(j) = \frac{1} {2}{({r_j} - {r^*})^2}{I_2} + \frac{1}
{6}{({r_j} - {r^*})^3}{D_3} + \frac{1} {{24}}{({r_j} - {r^*})^4}{D_5}$,
${S_4}(j) = \frac{1} {6}{({r_j} - {r^*})^3}{C_3} + \frac{1} {{24}}{({r_j} -
{r^*})^4}{C_5}$ and ${S_5}(j) = \frac{1} {{24}}{({r_j} - {r^*})^4}{I_2}$.
At $r = r_{j+1} >r^*$
\begin{gather*}
  {S_6}(j - 1){\Gamma _1} + {S_6}(j){\Gamma _2} + {\Gamma _3} + {\Gamma _4} + {\Gamma _5} = {B_1} + ({r_{j + 1}} -
  {r^*}){B_3} + \frac{1}
{2}{({r_{j + 1}} - {r^*})^2}{B_6}\\
  {S_7}(j - 1){\Gamma _1} + {S_7}(j){\Gamma _2} + ({r_{j + 1}} - {r^*}){\Gamma _3} + ({r_{j + 2}} -
  {r^*}){\Gamma _4} + ({r_{j + 3}} - {r^*}){\Gamma _5} = \\ \frac{1}
{{{r^*}}}{I_2} + ({r_{j + 1}} - {r^*}){B_2} + \frac{1}
{2}{({r_{j + 1}} - {r^*})^2}{B_5}\\
  {S_8}(j - 1){\Gamma _1} + {S_8}(j){\Gamma _2} + \frac{1}
{2}{({r_{j + 1}} - {r^*})^2}{\Gamma _3} + \frac{1} {2}{({r_{j + 2}} -
{r^*})^2}{\Gamma _4} + \frac{1} {2}{({r_{j + 3}} - {r^*})^2}{\Gamma _5} =
\\{I_2} + \frac{{({r_{j + 1}} - {r^*})}} {{{r^*}}}{I_2} + \frac{1}
{2}{({r_{j + 1}} - {r^*})^2}{B_4} \\
  {S_9}(j - 1){\Gamma _1} + {S_9}(j){\Gamma _2} + \frac{1}
{6}{({r_{j + 1}} - {r^*})^3}{\Gamma _3} + \frac{1} {6}{({r_{j + 2}} -
{r^*})^3}{\Gamma _4} + \frac{1} {6}{({r_{j + 3}} - {r^*})^3}{\Gamma _5} =\\
({r_{j + 1}} - {r^*}){I_2} + \frac{1}
{{2{r^*}}}{({r_{j + 1}} - {r^*})^2}{I_2}\\
  {S_{10}}(j - 1){\Gamma _1} + {S_{10}}(j){\Gamma _2} + \frac{1}
{{24}}{({r_{j + 1}} - {r^*})^4}{\Gamma _3} + \frac{1} {{24}}{({r_{j + 2}} -
{r^*})^4}{\Gamma _4} + \frac{1} {{24}}{({r_{j + 3}} - {r^*})^4}{\Gamma _5} =\\
\frac{1} {2}{({r_{j + 1}} - {r^*})^2}{I_2}
\end{gather*}
where ${S_6}(j) = {I_2} + ({r_j} - {r^*}){D_2} + \frac{1} {2}{({r_j} -
{r^*})^2}{F_2} + \frac{1} {6}{({r_j} - {r^*})^3}{F_4} + \frac{1} {{24}}{({r_j}
- {r^*})^4}{F_6}$, ${S_7}(j) = ({r_j} - {r^*}){C_2} + \frac{1} {2}{({r_j} -
{r^*})^2}{E_2} + \frac{1} {6}{({r_j} - {r^*})^3}{E_4} + \frac{1} {{24}}{({r_j}
- {r^*})^4}{E_6}$, ${S_8}(j) = \frac{1} {2}{({r_j} - {r^*})^2}{I_2} + \frac{1}
{6}{({r_j} - {r^*})^3}{D_4} + \frac{1} {{24}}{({r_j} - {r^*})^4}{D_6}$,
${S_9}(j) = \frac{1} {6}{({r_j} - {r^*})^3}{C_4} + \frac{1} {{24}}{({r_j} -
{r^*})^4}{C_6}$ and ${S_{10}}(j) = \frac{1} {{24}}{({r_j} - {r^*})^4}{I_2}$.
The systems at the points $r=r_{j-1}$ and $r=r_{j-2}$ are obtained in a similar
manner and will not be explicitly given here.

To demonstrate the convergence of the method we choose a fiber with multiple
layers that can support both TE and TM modes. Consider a fiber consisting of an
air core of radius $5\mu$m and a series of alternating layers of radii $1\mu$m
and refractive indices $n_1=2$ and $n_2=1$, respectively. The results for the
propagating constant for the TE and TM modes are shown in Table \ref{table.h4}.
\begin{table}[!htbp]
\centering
\begin{tabular}{ccccccc}
\hline
\raisebox{-1.50ex}[0cm][0cm]{Mode}&
\multicolumn{6}{c}{Propagation constant}  \\
\cline{2-7} & N=500 & N=1000 & N=2000 & N=4000 & N=10000 & N=20000  \\
\hline
TM & 0.52520655 & 0.52731320 & 0.52714039 & 0.52707988 & 0.52706043 & 0.52705446\\
4th order & 0.52732576 & 0.52707445 & 0.52705720 & 0.52705650 & 0.52705645 & 0.52705645\\
TE & 0.78582132 & 0.78580424 & 0.78588272 & 0.78590279 & 0.78590848 & 0.78590929\\
4th order& 0.78334621 & 0.78575156 & 0.78588122 & 0.78590670 & 0.78590941 & 0.78590954\\
\hline
\end{tabular}
\caption{The propagation constants for the
multi-layered fiber of the text.} \label{table.h4}
\end{table}
In Fig. \ref{h4.fig} the corresponding TE and TM modes are shown. The $m=1$
case is again described in the appendix.
\begin{figure}[!htbp]
    \centering
    \includegraphics[width=4.0in]{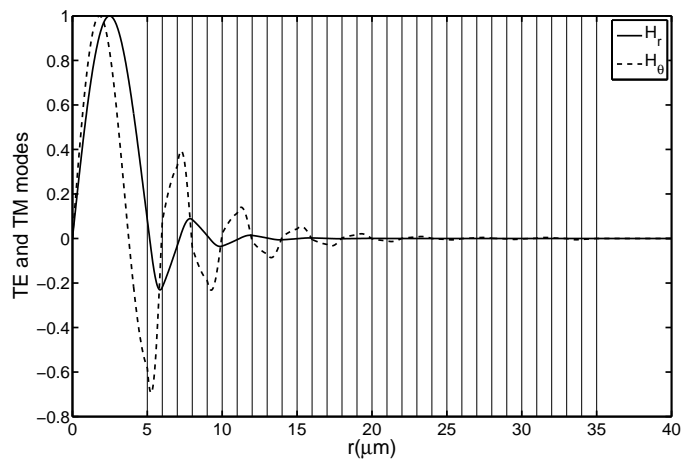}
    \caption{The TE and TM modes respectively corresponding to the propagation
    constants of Table \ref{table.h4}.}
    \label{h4.fig}
\end{figure}

One can go to even higher higher accuracy. In fact in Ref. \cite{zhou} the
authors have obtained sixteenth order accuracy. This, however, introduces a
fundamental limitation, namely the number of irregular points. In order for the
method to work, one must have a minimum number of points between interfaces.
When the structure of the fiber become more complicated it is obvious that this
is a serious limitation. We demonstrate how to overcome this in the following
section.
\section{Coordinate stretching transformation}
One of the problems one might encounter in this formalism is the total number
of points in a grid that must be used to have high accuracy results. The number
of grid points is directly proportional to the size of the matrix to be
diagonalized, thus more points means more computational time. When the width of
the fiber's core is considerably (orders of magnitude) larger than the width of
the layers, such as in the case of the fibers in Refs.
\cite{fink,nature,science2,johnson,kuriki}, one needs to use a sufficiently
small numerical step size to insure that there exist at least two points per
layer. If the number of points per interface is less than two, the assumption
used in the derivation of the method is not true, and the method fails. This is
because the method assumes that there is only one interface for each set of
three grid points. Because the reflecting layers are small, this means that a
small step size must be used. If one uses a uniform mesh, this means that in
the core region one is using a step size that is much smaller than necessary,
i.e., computational effort is being made unnecessarily.

To avoid this problem, we introduce a coordinate stretching by using a new
independent variable $\rho$ such that
\[
\rho  = \left\{ {\begin{array}{cc}
   {r,\quad r < R^* }  \\
   {R^*  + \sigma (r - R^* ),\quad r \geq R^* }
 \end{array} } \right.
\]
Here $R^*$  is the location of an additional artificial layer placed
arbitrarily inside the core of the fiber and $\sigma$ is a stretching parameter
such that $\sigma>1$. This transformation increases the effective width between
the layers relative to the core (by a factor of $\sigma$). We will use
uniformly spaced points in terms of $\rho$, which is equivalent to using a
smaller step size in the core and a larger step size in the layers when
measured in terms of $r$. Under this transformation, Eqs. (\ref{helm.vec})
become
\begin{subequations}
\begin{gather}
  {\bm{H}}_{\rho \rho }  + \frac{1}
{\rho }{\bm{H}}_\rho   + B{\bm{H}} = \beta^2 {\bm{H}},\quad \rho  < R^* \\
  \sigma ^2 {\bm{H}}_{\rho \rho }  + \frac{{\sigma ^2 }}
{{\rho  + (\sigma  - 1)R^* }}{\bm{H}}_\rho   + B{\bm{H}} = \beta^2
{\bm{H}},\quad \rho  \geq R^*
\end{gather}\label{helm.stretch}
\end{subequations}
At the point $\rho=R^*$ we impose the additional artificial jump conditions
\[{\bm{H}}^ +   = {\bm{H}}^ -  ,\quad{\bm{H}}_r^ +   = {\bm{H}}_r^
-
\]
so that, in terms of the new coordinate, both fields are continuous and their
derivatives satisfy
\[
\sigma {\bm{H}}_\rho ^ +   = {\bm{H}}_\rho ^ -
\]
It follows then, from Eq. (\ref{helm.vec}) and the above jump conditions that
the second derivatives satisfy
\[
\sigma ^2 {\bm{H}}_{\rho \rho }^ +   = {\bm{H}}_{\rho \rho }^ -
\]
where the $(-)$ and $(+)$ regions are on the left and right of the
discontinuity, respectively.  At the real layers, including the core, Eqs.
(\ref{cont.vec}) apply, with appropriate changes, namely
\[
  r = \frac{{\rho  - R^* }}
{\sigma } + R^*,\quad
  \frac{d}
{{dr}} = \sigma \frac{d} {{d\rho }}
\]
The finite difference approximation in matrix form at all points is again
\[
\Gamma _1 {\bm{H}}_{i - 1}  + \Gamma _2 {\bm{H}}_i  + \Gamma _3
{\bm{H}}_{i + 1}  = \beta^2 {\bm{H}}_i
\]
where the $\Gamma$'s are $2\times 2$ matrices. As always, we need to expand
$\bm{H}_{i-1}$, $\bm{H}_i$ and $\bm{H}_{i+1}$ around the points of the grid
that include the discontinuity. Assuming that $r_j\leq R^*<r_{j+1}$, the
systems that determine the finite difference coefficients at these regions then
satisfy the following systems

At $\rho_j\leq R^*$
\begin{gather*}
  \Gamma _1  + \Gamma _2  + \Gamma _3  = B \\
  (\rho _{j - 1}  - R^* )\Gamma _1  + (\rho _j  - R^* )\Gamma _2  + \Gamma _3 (\rho _{j + 1}  - R^* )/\sigma  =
  \frac{1}
{{R^* }}I_2  \\
  \frac{1}
{2}(\rho _{j - 1}  - R^* )^2 \Gamma _1  + \frac{1} {2}(\rho _j  - R^* )^2
\Gamma _2  + \frac{1} {{2\sigma ^2 }}(\rho _{j + 1}  - R^* )^2 \Gamma _3  = I_2
\end{gather*}
At $\rho_{j+1}>R^*$
\begin{gather*}
  \Gamma _1  + \Gamma _2  + \Gamma _3  = B \\
  \sigma (\rho _j  - R^* )\Gamma _1  + (\rho _{j + 1}  - R^* )\Gamma _2  + \Gamma _3 (\rho _{j + 2}  - R^* ) =
  \frac{\sigma }
{{R^* }}I_2  \\
  \frac{{\sigma ^2 }}
{2}(\rho _j  - R^* )^2 \Gamma _1  + \frac{1} {2}(\rho _{j + 1}  - R^* )^2
\Gamma _2  + \frac{1} {2}(\rho _{j + 2}  - R^* )^2 \Gamma _3  = \sigma ^2 I_2
\end{gather*}
The finite difference coefficients for the rest points are determined as usual
to be, at $\rho_j\leq \rho^*$
\begin{gather*}
  \Gamma _1  + \Gamma _2  + \Gamma _3 \left[ {I_2  + (\rho _{j + 1}  - \rho^* )D + \frac{1}
{2}(\rho _{j + 1}  - \rho^* )^2 F} \right] = B^ -   \\
  (\rho _{j - 1}  - \rho^* )\Gamma _1  + (\rho _j  - \rho^* )\Gamma _2  + \Gamma _3 \left[ {(\rho _{j + 1}  - \rho^* )C 
  +
  \frac{1}
{2}(\rho _{j + 1}  - \rho^* )^2 E} \right] = \frac{\sigma }
{{\rho^* }}I_2  \\
  \frac{1}
{2}(\rho _{j - 1}  - \rho^* )^2 \Gamma _1  + \frac{1} {2}(\rho _j  - \rho^* )^2
\Gamma _2  + \frac{1} {2}(\rho _{j + 1}  - \rho^* )^2 \Gamma _3  = \sigma ^2
I_2
\end{gather*}
and $\rho_j>\rho^*$
\begin{gather*}
  \Gamma _1 \left[ {I_2  - (\rho _j  - \rho^* )C^{ - 1} D + \frac{1}
{2}(\rho _j  - \rho^* )^2 F_2 } \right] + \Gamma _2  + \Gamma _3  = B^ +   \\
  \Gamma _1 \left[ {(\rho _j  - \rho^* )C^{ - 1}  + \frac{1}
{2}(\rho _j  - \rho^* )^2 E_2 } \right] + (\rho _{j + 1}  - \rho^* )\Gamma _2 +
\Gamma _3 (\rho _{j + 2}  - \rho^* ) = \frac{\sigma }
{{\rho^* }}I_2  \\
  \frac{1}
{2}(\rho _j  - \rho^* )^2 \Gamma _1  + \frac{1} {2}(\rho _{j + 1}  - \rho^* )^2
\Gamma _2  + \frac{1} {2}(\rho _{j + 2}  - \rho^* )^2 \Gamma _3  = \sigma ^2
I_2
\end{gather*}
where the matrices $E$ and $F$ are defined in section \ref{iim.coupled} and we
need to introduce the matrices
\[
E_2  = \left(\begin{array}{cc}
   0 & 0  \\
   0 & (1 - n_1^2 /n_2^2)/\rho^*
 \end{array} \right),\quad
 F_2  = \left( \begin{array}{cc}
   k^2 (n_2^2  - n_1^2 ) & 0  \\
   m(1 - n_1^2 /n_2^2)/\rho^*& k^2 (n_2^2  - n_1^2 ) -(1 - n_1^2 /n_2^2 )/\rho^{*2}
 \end{array} \right)
\]

\begin{figure}[!htbp]
    \centering
    \includegraphics[width=4in]{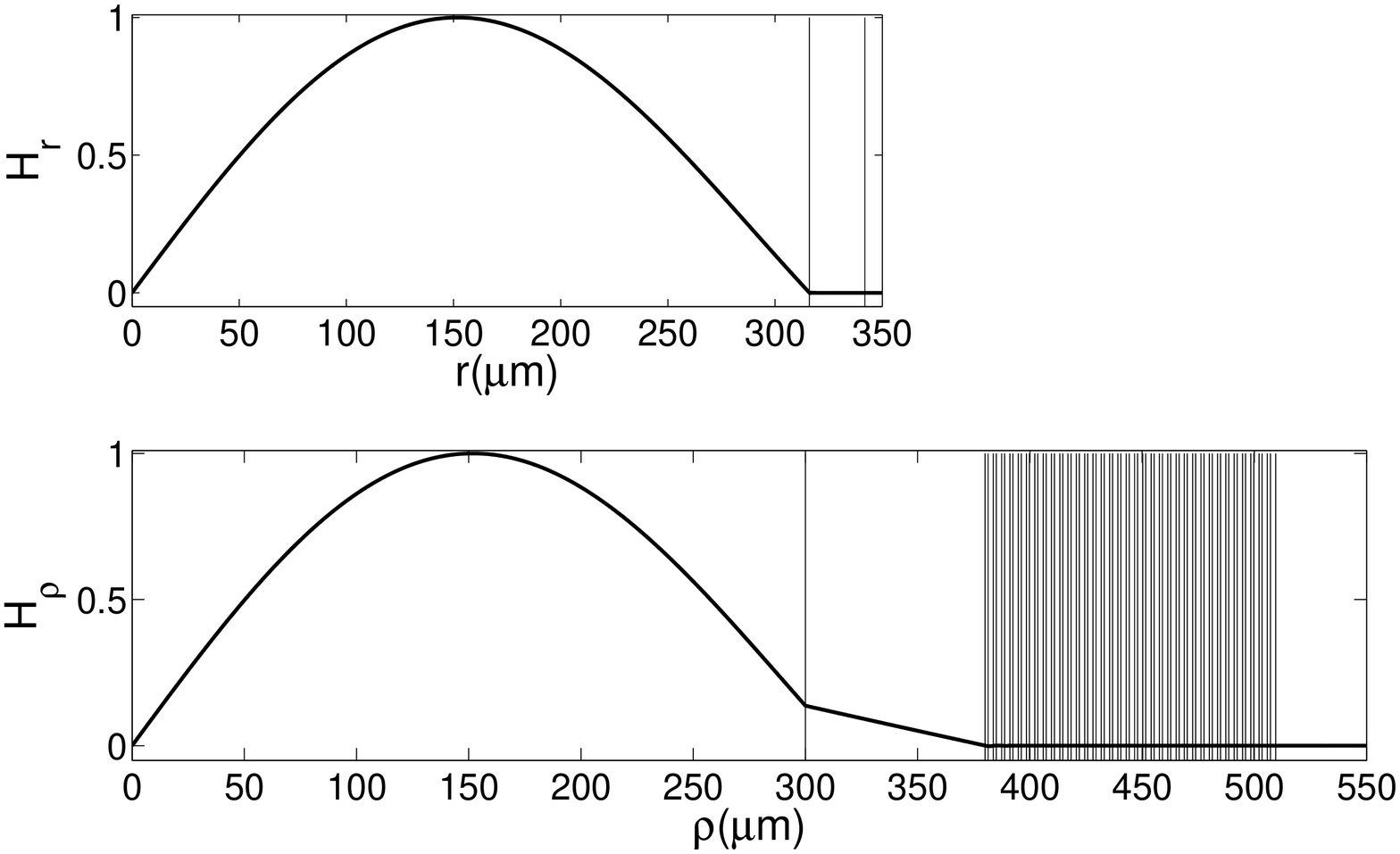}
    \includegraphics[width=4in]{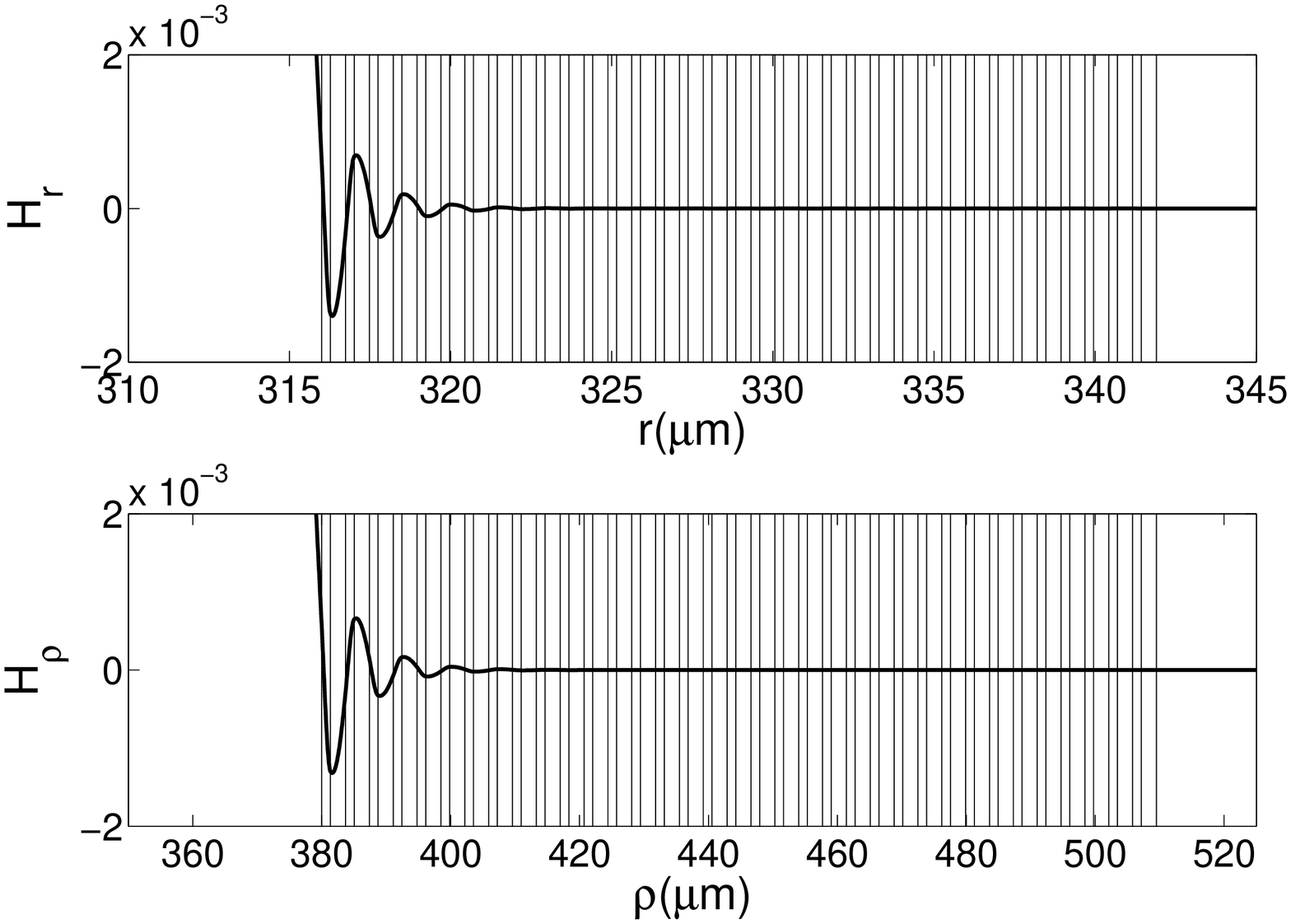}
    \caption{The TE mode of an Omniguide fiber using the original IIM and the coordinate stretching
    transformation. In the first figure only the first and last layers are plotted.}
    \label{stretch.fig}
\end{figure}

Using the coordinate stretching more demanding structures can be analyzed using
the same number of points. Consider, for example, the fiber described in Refs.
\cite{kuriki,horikis}. This fiber consists of a hollow-core $316\,\mu$m in
radius surrounded by 70 alternating layers of $\mathrm{As}_2\mathrm{Se}_3$
$0.27\,\mu$m thick, and polyether imide $0.47\,\mu$m thick. The fundamental
photonic bandgap is centered at $\lambda=2.28\,\mu$m and the refractive indices
of the layers are $n_1=2.8$ and $n_2=1.55+1.0{\times}10^{-4}\,{i}$,
respectively. The effective indices are found to be
$0.99999032+1.4{\times}10^{-12}\,i$ and $0.99999025+7.1{\times}10^{-11}\,i$ for
the TE and TM modes, respectively. These results were obtained for $20000$
points. Now consider the coordinate stretching transformation:
\[
\rho  = \left\{ {\begin{array}{cc}
   {r,\quad r < 300\mu m }  \\
   {300\mu m  + 5 (r - 300\mu m ),\quad r \geq 300\mu m }
 \end{array} } \right.
\]
For $5000$ points the IIM in its original formulation fails since the three
points per interface fails. However, we can get results of good accuracy using
the above transformation. Indeed, for $5000$ points the results are for the
effective index $0.99999030+1.39\times 10^{-12}\,i$ and $0.99999021+6.9\times
10^{-11}\,i$ for the TE and TM modes, respectively. In Fig. \ref{stretch.fig}
one can see and compare the TE modes obtained with and without the coordinate
stretching.
It is apparent that part from the added interface the fields are identical.
This transformation proves very useful for demanding structures such as this
one.
\section{Conclusions}
We presented a numerical method based on the immersed interface method that can
be used to obtain the propagation modes of circularly symmetric Bragg fibers
with arbitrary index profiles.  In its original formulation the method is
second order accurate and was applied to boundary value problems with
discontinuous and/or singular coefficients. We extended this method to matrix
eigenvalue problems and to higher accuracy. Cumbersome integrations or finding
roots of nontrivial functions, such as Bessel functions, are avoided and
computational time is minimized without sacrificing accuracy. All modes can be
determined and excellent results are achieved even for fibers with complicated
structure. Even when the geometry of the fiber is rather demanding, as in
Omniguide fibers, a coordinate stretching can be applied to keep computation
time to a minimum.
\section*{Acknowledgements}
I am very grateful to W.L. Kath for his help and support concerning this work.
\appendix

\section{Finite difference approximations for first and second derivatives}
For the second order scheme
\begin{eqnarray*}
\frac{dH}{dr} &=& \frac{H_{i+1}-H_{i-1}}{2h}\\
\frac{d^2H}{dr^2} &=& \frac{H_{i+1}-2H_{i}+H_{i-1}}{h^2}
\end{eqnarray*}
and for the fourth order case
\begin{eqnarray*}
\frac{dH}{dr} &=& \frac{-H_{i+2}+8H_{i+1}-8H_{i-1}+H_{i-2}}{12h}\\
\frac{d^2H}{dr^2} &=& \frac{-H_{i+2}+16H_{i+1}-30H_{i}+16H_{i-1}-H_{i-2}}{12h^2}
\end{eqnarray*}

\section{Boundary conditions for higher-order differences}
As usual, special care must be given to the end points. High-order methods
require an extrapolation scheme to determine the differencing coefficients at
$r_2$ and $r_{N-1}$. Given the boundary conditions $\bm{H}(r_1)=\bm{H}_1=0$ and
$\bm{H}(r_N)=\bm{H}_N=0$ we use the extrapolation
\begin{gather*}
  {{\bm{H}}_2} =  - {{\bm{H}}_6} + 4{{\bm{H}}_5} - 6{{\bm{H}}_4} + 4{{\bm{H}}_3} \\
  {{\bm{H}}_{N - 1}} = 4{{\bm{H}}_{N - 2}} - 6{{\bm{H}}_{N - 3}} + 4{{\bm{H}}_{N - 4}} - {{\bm{H}}_{N - 5}}
\end{gather*}
to accommodate for these extra points.

\section{Special case boundary conditions ($\boldsymbol{m=1}$)}

All of the cases considered had the fields tend to zero at the boundaries,
namely at $r=0$ and $r\rightarrow\infty$. However, this is not the case for the
calculation of the $HE_{11}$ mode (our $m=1$ case). Indeed, recall that the
boundary conditions for this case are
\[
{H_{r}} + {H_{\theta}} = 0\quad \text{and}\quad \frac{d}{{dr}}\left( {{H_{r}} -
{H_{\theta}}} \right) = 0
\]
at $r=0$. To accommodate this we need to alter the finite difference scheme to
include the new boundary conditions.

\subsection{The second order correction}

At the first point where $i=1$ Eqs. \eqref{te.new} become
\begin{subequations}
\begin{gather}
\gamma _1 H_{r,0}  + \gamma _2 H_{r,1}  + \gamma _3 H_{r,2}  + \Delta H_{\theta ,1}  = \beta ^2 H_{r,1}\\
  \delta _1 H_{\theta ,0}  + \delta _2 H_{\theta ,1}  + \delta _3 H_{\theta ,2}  + \Gamma H_{r,1}  = \beta ^2
  H_{\theta ,1}
\end{gather}
\label{te.new.app}
\end{subequations}
and the boundary conditions become under a second order approximation of the
first derivative
\[
H_{r,1}+H_{\theta,1}=0 \quad \text{and}\quad H_{r,2}-H_{r,0}=H_{\theta,2}-H_{\theta,0}
\]
Note that all the coefficients are known (we are not at an interface) and we
only need to eliminate the two $H_{r,0}$ and $H_{\theta,0}$  terms.

Adding Eqs. \eqref{te.new.app} and taking the limit as $r\rightarrow 0$ gives
(recall $\gamma_1=\delta_1$, $\gamma_2=\delta_2$, $\gamma_3=\delta_3$ and
$\Gamma=\Delta$ as clearly seen from their definition below Eqs.
\eqref{te.new})
\[
H_{r,2}+H_{r,0}=H_{\theta,2}+H_{\theta,0}
\]
which suggests that
\[
\bm{H}_{0}=\bm{H}_{2}
\]

\subsection{The fourth order correction}

At the first point where $i=1$ we write the system as
\begin{subequations}
\begin{gather}
\gamma _1 H_{r,-1}  + \gamma _2 H_{r,0}  + \gamma _3 H_{r,1}  + \gamma _4 H_{r,2}  + \gamma_5 H_{r,3}  + \Delta
H_{\theta ,1}  = \beta ^2 H_{r,1}\\
\delta _1 H_{\theta ,-1}  + \delta _2 H_{\theta ,0}  + \delta _3 H_{\theta ,1}
+ \delta _4 H_{\theta ,2}  + \delta _5 H_{\theta ,3}  + \Gamma H_{r,1}  =
\beta^2 H_{\theta ,1}
\end{gather}
\label{te.new.h4}
\end{subequations}
where the definitions for the coefficients follow from Eq. \eqref{fd.h4}. The
steps here follow the procedure for the second order results. Thus, from the
definition of the first derivative at fourth order for $i=1$ we get
\[
H_{r,-1}-8H_{r,0}+8H_{r,2}-H_{r,3}=H_{\theta,-1}-8H_{\theta,0}+8H_{\theta,2}-H_{\theta,3}
\]
As above adding Eqs. \eqref{te.new.h4} and taking the limit $r\rightarrow 0$
gives
\begin{gather}
\frac{1}{12}(H_{r,-1}+H_{\theta,-1})-\frac{2}{3}(H_{r,0}+H_{\theta,0})+\frac{2}{3}(H_{r,2}+H_{\theta,2})
-\frac{1}{12}(H_{r,3}+H_{\theta,3})=0 \label{app.h4.2}
\end{gather}
However, we also need to extrapolate for the point $i=-1$, namely
\begin{gather}
\bm{H}_{-1}= 4 \bm{H}_0 - 6 \bm{H}_1 + 4 \bm{H}_2 - \bm{H}_3 \label{app.h4.3}
\end{gather}
Note that while the point $i=0$ was before ignored this is not the case here.
Using these equations we can now eliminate the point $i=0$ as follows
\begin{gather*}
\bm{H}_{0} = -\frac{3}{2} \bm{H}_{1} + 3 \bm{H}_{2} - \frac{1}{2} \bm{H}_{3}
\end{gather*}

\section*{References}
\bibliographystyle{elsarticle-num}

\end{document}